\documentclass[
  aip,
  amsmath,
  amssymb,
  preprint,
  groupedaddress,
  ]{revtex4-1}

\usepackage[pdftex]{graphicx}
\usepackage{dcolumn}% Align table columns on decimal point
\usepackage{bm}% bold math
\usepackage{comment}

\usepackage[utf8]{inputenc}
\usepackage[T1]{fontenc}
\usepackage{mathptmx}
\usepackage{etoolbox}

\makeatletter
\def\@email#1#2{%
 \endgroup
 \patchcmd{\titleblock@produce}
  {\frontmatter@RRAPformat}
  {\frontmatter@RRAPformat{\produce@RRAP{*#1\href{mailto:#2}{#2}}}\frontmatter@RRAPformat}
  {}{}
}%
\makeatother

\begin{document}

%\preprint{}

%\title{Dynamic nuclear polarization of $^{27}$Al and $^{139}$La in Nd$^{3+}$LaAlO$_3$ crystals grown by optical Floating-Zone method}
%\title{Crystal growth recipe of Nd$^{3+}$:LaAlO$_3$ crystals based on optical floating-zone method for dynamic nuclear polarization of $^{139}$La and $^{27}$Al}
\title{Combination of crystal growth with optical floating zone and evaluation of Nd$^{3+}$:LaAlO$_{3}$ crystals with the dynamic nuclear polarization of 
$^{139}$La and $^{27}$Al}

\author{Kohei Ishizaki$^1$}
\email{ishizaki@phi.phys.nagoya-u.ac.jp}
\author{Ikuo Ide$^1$}
\author{Masaki Fujita$^2$}
\author{Hiroki Hotta$^1$}
\author{Yuki Ito$^1$}
\author{Masataka Iinuma$^3$}
\author{Yoichi Ikeda$^2$}
\author{Takahiro Iwata$^4$}
\author{Masaaki Kitaguchi$^5$}
\author{Hideki Kohri$^{6,1}$}
\author{Taku Matsushita$^1$}
\author{Daisuke Miura$^4$}
\author{Yoshiyuki Miyachi$^4$}
\author{Hirohiko M. Shimizu$^1$}
\author{Masaru Yosoi$^6$}

\affiliation{
    $^1$Nagoya University, Furocho, Chikusa, Nagoya, 464-8602, Japan \\
    $^2$Institute for Materials Research, Tohoku University, Sendai 980-8577, Japan \\
    $^3$Hiroshima University, Kagamiyama, Higashi-hiroshima, 739-8527, Japan \\
    $^4$Yamagata University, Koshirakawa, Yamagata 990-8560, Japan \\
    $^5$KMI, Nagoya University, Furocho, Chikusa, Nagoya, 464-8602, Japan \\
    $^6$Research Center for Nuclear Physics, Osaka University, Ibaraki, Osaka, 567-0047, Japan
    } %\\

\date{\today}

%\begin{abstract}
%Polarized target research often requires special experiments technique in terms of sample preparation and cryogenics. We have established a fundamental way for growing the LaAlO$_3$ crystals based on the optical floating zone method with halogen lamps and examining the characteristics as the DNP target for each samples. We demonstrate the effectiveness of our research method by successfully preparing two samples with some La replaced with different amounts of the Nd : 0.05 mol\%, 0.01 mol\% and achieving the $3.5\pm0.3$ and $13 \pm 3$ enhancements for each Nd : 0.05 mol\% and 0.01 mol\% samples.
%\end{abstract}

\begin{abstract}
%Dynamic Nuclear Polarization (DNP), which is known as a technique for polarizing nuclear spins, is used for a polarized target in particle and nuclear physics experiments and is mostly quite sensitive to characteristics of target material. For realizing a polarized Lanthanum ($^{139}$La) target, we have established a fundamental way of growing single crystals of Nd$^{3+}$:LaAlO$_3$ based on the optical floating zone method with halogen lamps and examined quantities indicating applicability as the DNP target for prepared samples. The effectiveness of our method has been confirmed by observing enhancement of NMR signals of $^{139}$La and $^{27}$Al in two samples, which is eventually $3.5\pm0.3$ and $13 \pm 3$ for the concentration of Nd ions with 0.05 mol\% and 0.01 mol\%, respectively. The present technique of combining the crystal growth and the DNP is effectively applicable to an optimization of amount of the Nd ions, which is essential for developing a polarized $^{139}$La target for practical use. 
Producing a polarized lanthanum (La) target with high polarization 
and long relaxation time is crucial for realizing 
time-reversal violation experiments using polarized neutron beams. 
We use a LaAlO$_{3}$ crystal doped with a small amount of Nd$^{3+}$ 
ions for the polarized lanthanum target. 
Optimizing the amount of Nd$^{3+}$ ions is considerably important because 
the achievable polarization and relaxation time 
strongly depend on this amount. 
We established a fundamental method to grow single crystals of 
Nd$^{3+}$:LaAlO$_3$ using an optical floating zone method that 
employs halogen lamps and evaluated the 
crystals with the dynamic nuclear polarization (DNP) method 
for polarizing nuclear spins. 
Two crystal samples were grown by ourselves and evaluated with the DNP at 1.3 K and 2.3 T for the first time except for the target materials of protons. 
The enhancement of NMR signals for $^{139}$La and $^{27}$Al was successfully observed, 
and the enhancement factors were eventually $3.5\pm0.3$ and $13 \pm 3$ for the samples 
with Nd$^{3+}$ ions of 0.05 and 0.01 mol\%, respectively. 
These enhancement factors correspond to absolute vector polarizations of 
0.27$\pm$0.02\% (Nd$^{3+}$ 0.05 mol\%) and 1.4$\pm$0.3\% (Nd$^{3+}$ 0.01 mol\%).  
Although the obtained polarizations are still low, they are acceptable 
as a first step. 
The combination scheme of the crystal growth and evaluation of the crystals 
is found to be effectively applicable for optimizing the amount of 
Nd$^{3+}$ ions for improving the performance of the polarized target. 
\end{abstract}

\pacs{}
%\keywords{}

\maketitle


\section{Introduction\label{sec:ontroduction}}
Wu et al. first found the violation of discrete symmetry in a nucleus via 
polarized $^{60}$Co experiments~\cite{Wu1957}. 
Similar parity non-conservation effects in nuclei were observed in the 
polarized neutron scattering with the unpolarized nuclei. 
Many types of nuclei, 
including lanthanum($^{139}$La), showed large asymmetry 
in the absorption cross sections of a polarized neutron. 
The parity-violating asymmetry was enhanced to approximately $10^2-10^6$ times 
larger than that of the proton-proton scattering~\cite{Mitchell2001}. 
The asymmetry can be explained by both the multi-body effect in the nucleus and the 
interference between the large parity-conserving scattering amplitude and tiny 
parity-nonconserving one~\cite{Mitchell2001}. 
Theoretical studies predicted similar enhancements in the time-reversal 
violating interaction for certain nuclei~\cite{Gudkov1992}. 
Recent experimental results showed that $^{139}$La is a promising target 
for exploring the enhancement-assisted 
T-violation effects~\cite{Okudaira2018, Yamamoto2020}.

One difficulty in the T-violation experiment is realizing the 
polarized $^{139}$La target. 
One possible method to achieve high polarization is dynamic nuclear polarization (DNP) 
with a LaAlO$_3$ crystal, where a small amount of La$^{3+}$ ions are replaced with 
Nd$^{3+}$ ions behaving as a paramagnetic dopant in the DNP\cite{Takahashi1993}. 
In the past, two DNP experiments reported the enhancement of the vector 
polarization of $^{139}$La in the Nd$^{3+}$:LaAlO$_3$ crystal with Nd 0.03 mol\%. 
The vector polarization $P_V$ is defined as 
\begin{equation}
  P_V = \frac{\sum_{m=-I}^{m=+I} mP_m}{I\sum_{m=-I}^{m=+I} P_m}, 
  \label{eq:vector_polarization}
\end{equation}
where $I$ represents a nuclear spin, $m$ represents a spin quantum number, 
$P_m$ represents the population on state $m$ of the target nuclear spins. 
Hereafter, the vector polarization is expressed as "polarization". 
The first experiment was conducted at approximately 1.5 K and 
2.3 T~\cite{Maekawa1995}, 
and the polarization of 0.2 for $^{139}$La was achieved. 
In another experiment, polarizations of 0.62 and 0.475 for $^{27}$Al and $^{139}$La 
in the Nd 0.03 mol\% crystal were obtained using a dilution refrigerator 
at 2.3 T~\cite{Hautle2000}. 
These results show the potentiality of the Nd$^{3+}$:LaAlO$_3$ crystal 
as the polarized target. 

In the DNP, a dominant relaxation process of nuclear spins is caused by 
the magnetic fluctuation of paramagnetic ions, which are coupled to nuclear spins 
via the dipole-dipole interaction. 
Hence, such relaxation effects can be suppressed by reducing 
the Nd$^{3+}$ amount. 
However, it reduces the efficiency of polarization transfer from the electron 
polarization to nuclear spins.
For increasing polarization, it is necessary to determine the 
optimal Nd$^{3+}$ amount 
by studying the relationship between Nd$^{3+}$ concentration and 
polarization enhancement. 
%Other possible problems are the amount of unintended paramagnetic impurities and crystal defects, 
%which affect the DNP. 
Establishing a fundamental method for growing the crystal 
samples by ourselves is preferred because it allows us to flexibly prepare various 
samples with different Nd$^{3+}$ amounts. 

LaAlO$_3$ crystals are usually produced via the Czochralski 
method~\cite{Fay1967,Zeng2004}. 
The Bridgman method is also applicable to the growth of 
the LaAlO$_3$ crystals~\cite{Fahey1993}. 
These methods are suited for growing large crystals; however, they are too costly 
to flexibly 
prepare several appropriate crystals required for studying fundamental properties. 
In addition, the precise control of unexpected impurities and 
the required Nd$^{3+}$ 
amounts are difficult because some procedures with a crucible are 
unavoidable in these methods.
In contrast, an optical floating zone (FZ) method is applicable for growing 
a good-sized sample at a low cost. 
The properties of the LaAlO$_3$ crystals are low optical absorption and 
high melting point (2100 $^{\circ}$C), and therefore, 
to the best of our knowledge regarding the optical FZ method, 
there are no reports of growing LaAlO$_3$ crystals except 
for melting by xenon lamps~\cite{Suzuki2005, Inagaki2007, Bednorz1984}. 
Instead of the xenon lamps, we attempted to grow a crystal without any dopant and 
a crystal with Nd$^{3+}$ 0.05 mol\% using halogen lamps~\cite{Ishizaki2020} 
although an achievable temperature was comparable to the melting point. 
One benefit of the latter method is that a generalized apparatus is available, 
which is more versatile and inexpensive than the former. 
The effectiveness of this approach can be verified by 
performing simple DNP experiments. 

In this paper, we present the first confirmation regarding the effectiveness 
of the optical FZ method using halogen lamps and the first results of 
the DNP experiments of Nd$^{3+}$:LaAlO$_3$ crystals produced by ourselves 
with the DNP at 1.3 K and 2.3 T. 
%The prepared samples have two different amounts of Nd : 0.05 mol\% and 0.01 mol\%, 
%which are also different from the previous experiments. 
%The environment of the DNP experiments was at the temperature of 1.5 K and in 
%the magnetic field of approximately 2.3 T, similar to the past experiment~\cite{Maekawa1995}. 
%The latest status on the development of the polarized $^{139}$La target can be 
%referenced in our different literature~\cite{Ide2023}, which briefly reports 
%the large enhancement of the $^{139}$La with the frequency-optimized microwave. 
%On the other hand, here, we discuss details of combining techniques of the sample 
%preparation and the DNP experiments because they are indicators for examining 
%the feasibility of an efficient search for an optimal amount of the Nd ions, 
%which is an essential task for the development of the polarized La target.
The rest of the paper is organized as follows. 
In section 2, we focus on the importance and motivation of the combination 
scheme of the crystal growth and the evaluation with DNP. 
Section 3 explains the crystal growth with the FZ method using halogen lamps. 
Section 3 also shows the experimental results of the DNP experiments 
with two types of crystal samples prepared with the optical FZ method. 
%In section 4, we discuss how to improve the present way of the crystal growth 
%and evaluation of the crystals. 
Finally, section 4 summarizes the results and presents 
the future prospects. 

\section{Procedure of crystal growth and evaluation of crystals}

\begin{figure}[ht]
  \centering
  \includegraphics[keepaspectratio, scale=0.5]{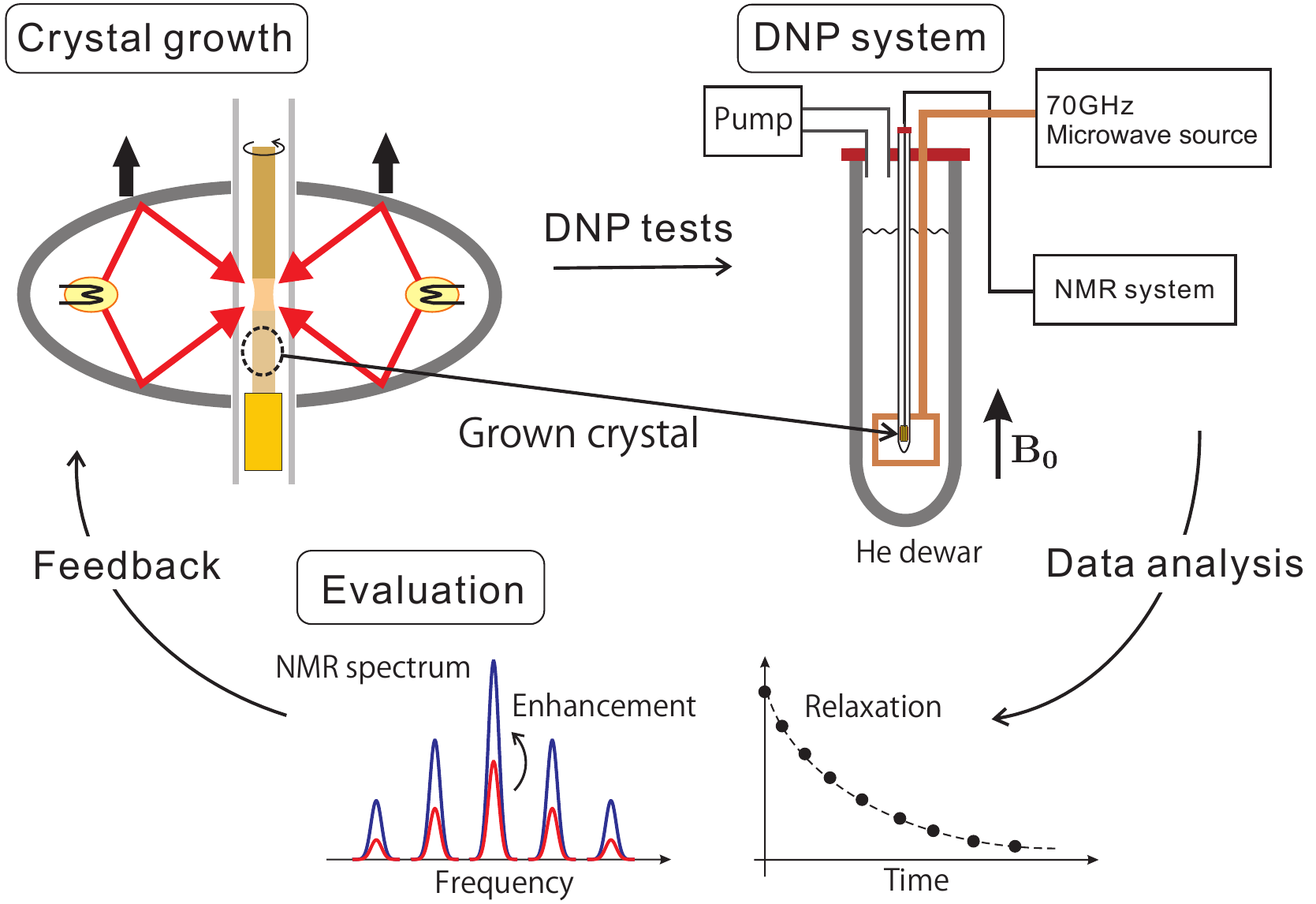}
  \caption{Combination scheme of the crystal growth and evaluation of crystals with 
           the DNP.}
  \label{fig:procedure}
\end{figure}

The achievable polarization and relaxation time are strongly dependent on 
the amount of Nd$^{3+}$ ions in a LaAlO$_{3}$ crystal, and therefore, 
many cycles of crystal 
growth and evaluations are required to optimize the amount of 
Nd$^{3+}$ ions. 
Figure~\ref{fig:procedure} shows the procedures of a cycle. 
Some LaAlO$_{3}$ crystals with different amounts of Nd$^{3+}$ ions 
are grown in the first step. 
The crystals are cooled at $\sim$1.5 K, and DNP is performed using 
microwaves at a magnetic field of a few Tesla in the second step. 
The polarizations and relaxation times are obtained by data analysis 
in the third step. 
Feedback from the results of the data analysis improves the next cycle, 
and the cycles are repeated until the optimal amount of the Nd$^{3+}$ ions 
is determined. 
A fundamental approach for growing the crystals needs to be established 
to perform a cycle, which is sufficiently effective for the DNP, 
with precise control of the Nd amount in the range of 0.01 mol\%. 
Furthermore, it is desirable for the DNP system to be simple and easy to use 
for implementing flexible and rapid DNP tests. 
The first issue for moving the cycle into action is to check the control of the 
Nd amount in the crystal growth. 
Therefore, it is necessary to grow two crystals with different Nd amount 
and to confirm whether their difference can be observed 
in the DNP evaluation. 

\section{Crystal growth using floating zone method\label{sec:crystal_growth}}
\subsection{Structure of LaAlO$_3$ crystals}

The crystal structure of LaAlO$_3$ and its dependence on the temperature have been 
studied using neutron powder diffraction~\cite{Hayward2005}. 
Above 813 K, the crystal structure is perovskite, whereas a distorted-perovskite 
structure appears below 813 K, and the unit cell only has a threefold 
rotation ($C_3$) axis. 
Even in the structure below 813 K, all La and Al sites are magnetically equivalent 
and have uniaxial magnetic symmetry, respectively. 
The distortion is so tiny that LaAlO$_3$ crystals spontaneously form twin domains 
in the structural phase transition. 
All domains in the crystal are assorted into four groups according to the direction 
of the $C_3$ axis. 
In the DNP experiments, the crystal samples must be aligned so that 
the $C_{3}$ axis of one domain becomes parallel to the external magnetic field. 
In this paper, the domain with the $C_3$ axis parallel to the external magnetic field 
is referred to as the "primary domain", and the other domains are 
the "secondary domain" .

\subsection{Crystal growth}

\begin{figure}[ht]
  \centering
  \includegraphics[keepaspectratio, scale=0.3]{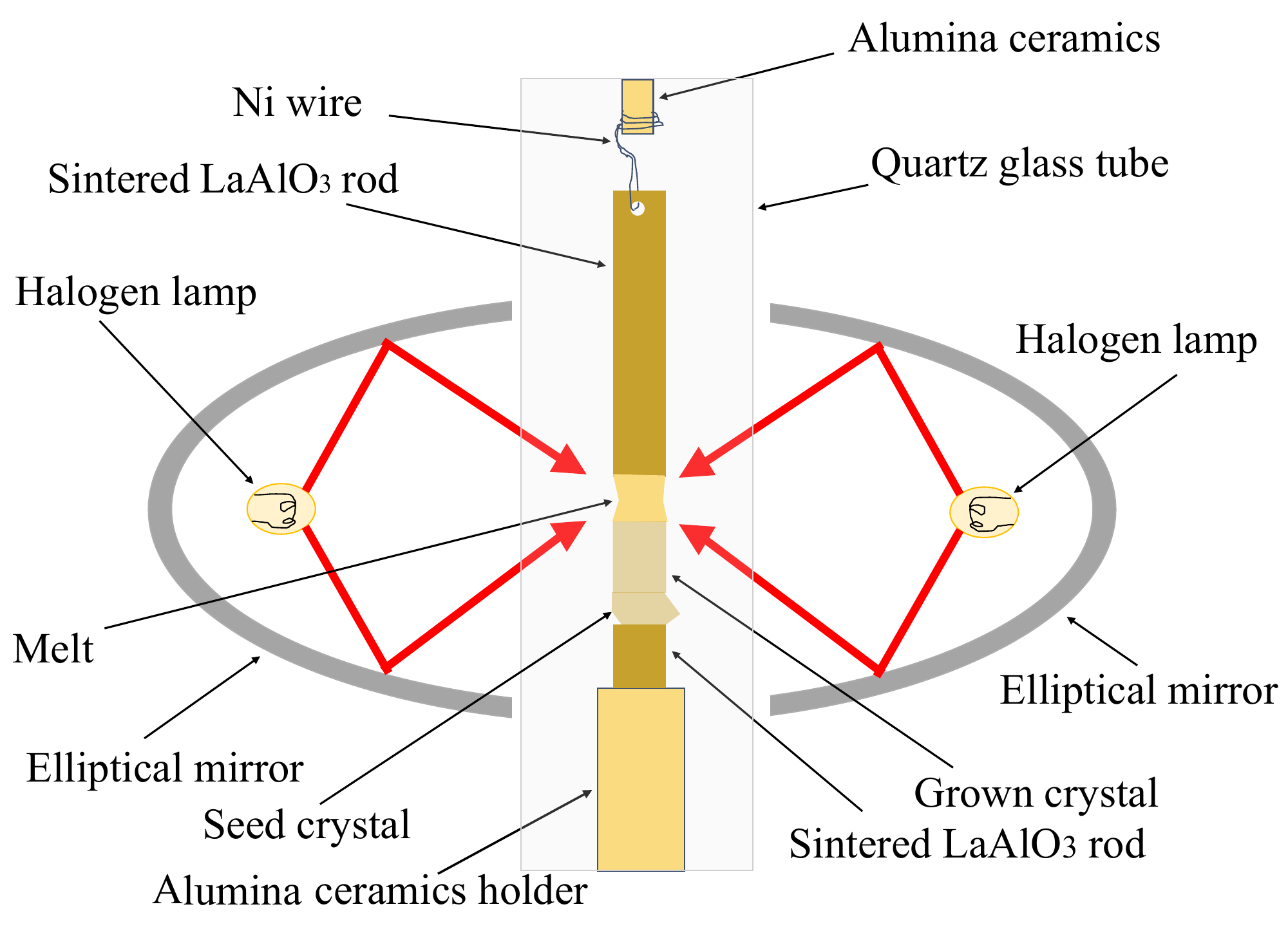}
  \caption{Schematic of the FZ apparatus used for crystal growth. 
           The FZ apparatus is equipped with four halogen lamps, but only two mirrors are 
            drawn for clarity.}
  \label{fig:fz}
\end{figure}

The details of our crystal growth method using halogen lamps have been explained 
elsewhere~\cite{Ishizaki2020}. 
Here, we present a brief summary. 
The raw materials were La(OH)$_3$ (4N), Al$_2$O$_3$ (4N), and Nd$_2$O$_3$ 
(3N). 
Each powder of LaAlO$_3$ and La$_{0.99}$Nd$_{0.01}$AlO$_3$ was synthesized by mixing 
the raw materials according to the stoichiometric ratio and calcining at 1400 $^{\circ}$C 
for over 8 h. 
The powder of Nd$^{3+}$:LaAlO$_3$ with Nd 0.05 mol\% or 0.01 mol\% was prepared 
by mixing LaAlO$_3$ and La$_{0.99}$Nd$_{0.01}$AlO$_3$. 
The feed and seed rods were shaped by cold isostatic pressing and 
sintering them at 1400 $^{\circ}$C for over 8 h. 
The seed crystal was placed on the seed rod to grow the crystal along the $C_3$ axis. 
As shown in Fig.~\ref{fig:fz}, the feed rod and seed crystal were melded by 
converging the infrared ray in the FZ furnace (Crystal Systems Co.), which 
consisted of four 1 kW halogen lamps. 
The optimal condition for growing was 10 mm/h in the air. 
Samples were successfully grown to lengths ranging from 10-20 mm and 
diameters of $\sim$5 mm, which is suitable for 
developing polarized targets.

%%%%%%%%%%%%%%%%%%%%%%%%%%%

For the DNP experiments, we prepared two crystal samples with 
Nd 0.05 and 0.01 mol\%. 
The former was dark brown and the latter was light brown.
Although the reason of such a difference of color is not specified 
yet, the problem is expected to be resolved by repeating 
the cycle shown in Fig.~\ref{fig:procedure}. 
At this time, therefore, it is important to directly compare 
the results of the DNP between these crystals. 

%For the DNP experiments, we prepared two crystal samples with Nd 0.05 and 
%0.01 mol\%. 
%The former was dark brown and the latter was light brown. 
%Such a difference of color plausibly reflects
%the distinction of the Nd concentration~\cite{Ishizaki2020}. 
%At this time, the results of the DNP can be directly compared 
%between these crystals. 
%The impurities can also enter the materials in the above process because of 
%accidental contamination of the used apparatus. 
%It is reasonable to consider that the colors are not derived from the Nd$^{3+}$ %ions because grown samples without doping the Nd$^{3+}$ ions were sometimes %colored brown~\cite{Ishizaki2020}. 

\subsection{Selection of grown crystals with EPR and NMR \label{sec:crystal_eval}}

We confirmed that the grown samples had a single crystal and that the growth direction 
corresponded to the $C_3$ axis in the cubic phase using the X-ray Laue technique. 
In the next step, the electron paramagnetic resonance (EPR) of Nd$^{3+}$ ions 
in the 0.05 mol\% sample was measured at 10 K using the CW-EPR with 
a 9.5 GHz microwave. 
The effective $g$-factor and EPR linewidth at full-width-half maximum in the primary 
domain were obtained as 2.12 and 6 G, respectively. 
These values were comparable to those in the LaAlO$_3$ sample used in the previous 
experiments~\cite{Takahashi1993,Ishizaki2020}. 
The linewidth is attributed to the super-hyperfine interaction 
between the $^{27}$Al nucleus and Nd$^{3+}$ ions~\cite{Takahashi1993}. 

%%%%%%%%%%%%%%%%%%%%%%%%%%%%%%

The EPR measurement revealed that the volume of the domain with the $C_3$ axis 
parallel to the growth direction was small in the sample with Nd 0.05 mol\%. 
We measured NMR spectra at 300 K and 6.9 T using a conventional 
pulse-NMR system. 
The direction of the $C_{3}$ axis of the largest domain was identified 
by changing orientation of the crystal. 
DNP experiments can be performed when the magnetic field is applied to the $C_{3}$ 
axis of the largest domain. 

%%%%%%%%%%%%%%%%%%%%%%%%%%%%%%%

The frequency of the transition $m\leftrightarrow m+1$ of the nuclear spin with 
$I ( > 1/2)$ in the uniaxial electrical field gradient such as $^{139}$La ($I=7/2$) 
and $^{27}$Al ($I=5/2$) in LaAlO$_3$ is expressed as 
\begin{equation}
  \nu_{m \leftrightarrow m+1} = \nu_L - \nu_Q \left( \frac{3\cos ^2 \theta -1}{2} \right) \left( m + \frac{1}{2} \right), 
  \label{eq:resonance_frequency}
\end{equation}
where $\nu_{L}$, $\nu_{Q}$, and $\theta$ represent 
the Larmor frequency of the nucleus, 
frequency shift parameter derived from the nuclear quadrupole interaction, and 
angle between the external magnetic field and symmetrical axis of the 
electric-field gradient, 
which corresponds to the $C_{3}$ axis in the domains~\cite{Hautle2000}. 
$\nu_Q$ depends on the sample temperature 
because the lattice parameters depend on the temperature. 
Further, each frequency shift parameter $\nu_Q$ for $^{27}$Al in LaAlO$_3$ is $+22$ kHz 
and $+30$ kHz at 300 K and 1.3 K, respectively, 
whereas that for $^{139}$La is $+495$ kHz and $+673$ kHz at 300 K and 1.3 K, 
respectively. 
The absolute value of $\nu_Q$ was measured in our experiments, and the sign was determined 
from the previous experiments~\cite{Hautle2000}.

The angle between the $C_3$ axis of the secondary domains and external magnetic field 
is approximately 70$^{\circ}$. 
The frequency shift induced by the nuclear quadrupole interaction in the secondary domains 
is one-third of that in the primary domain. 
In the NMR spectrum, the central frequency of each peak 
distinguishes between the primary or secondary domain 
using Eq.~(\ref{eq:resonance_frequency}). 
The evaluation of the area of peaks leads to the ratio of the domain's volume in the sample. 

\section{Evaluation of crystals with the DNP method}\label{sec:dnp}
\subsection{DNP apparatus}

\begin{figure}[ht]
    \centering
    \includegraphics[width=14.cm]{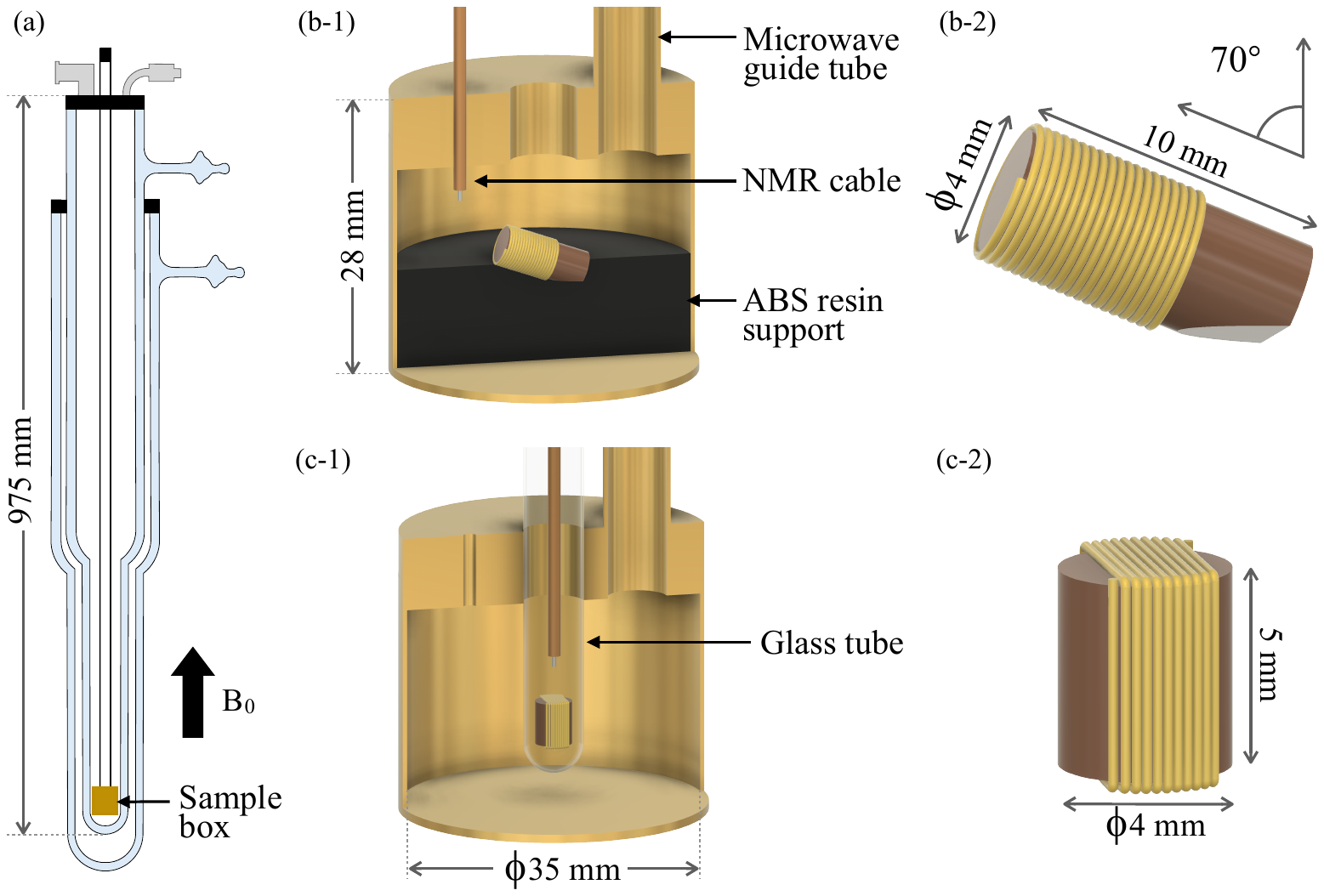}
    \caption{Apparatuses of two DNP experiments for crystal 
            samples with (a) liquid He dewar, (b-1) sample box of Nd 0.05 mol\%,  and (c-1) sample box of Nd 0.01 mol\%. 
            The NMR coils and LaAlO$_{3}$ crystal samples with 
            Nd (b-2) 0.05 and (c-2) 0.01 mol\%.}
    \label{fig:apparatus}
\end{figure}

Figure~\ref{fig:apparatus} shows the apparatuses of two DNP experiments 
for crystal samples with Nd 0.05 and 0.01 mol\%. 
The samples inside the sample box drawn in Fig. \ref{fig:apparatus} (b-1) or (c-1) were 
cooled down to 1.3 K by pumping liquid $^4$He in the glass dewar, as shown in Fig. 
\ref{fig:apparatus} (a). 
The space between the inner and outer layers of the dewar was maintained as a vacuum 
to ensure thermal insulation.  
The outer layer was cooled by liquid nitrogen to reduce the 
consumption of liquid $^4$He. 
The temperature was monitored with a ruthenium oxide thermometer and 
a platinum resistance thermometer attached to a brass box 
where the sample was installed. 

In the first DNP experiment, the sample with Nd 0.05 mol\% 
was fixed on the base plate made of ABS resin, which was produced 
with a 3D printer, as shown in Fig.~\ref{fig:apparatus} (b-1). 
We created a tiny cut in the sample as shown in Fig.~\ref{fig:apparatus} (b-2) 
to recognize the direction of the $C_{3}$ axis of the largest domain. 
The sample was fixed so that the tiny cut was in contact with the ABS resin, and 
the $C_3$ axis of the largest domain became parallel 
to the external magnetic field {\rm B$_{0}$}. 
The sample was directly cooled through thermal contact to 
the liquid $^4$He. 

In the second DNP experiment, we used a column-shaped sample with 
Nd 0.01 mol\%, which was 4 mm in diameter 
and 5 mm in length, as shown in Fig.~\ref{fig:apparatus} (c-1, c-2). 
The sample was aligned such that the growth direction was parallel to the external 
magnetic field {\rm B$_{0}$} and placed in the quartz tube filled with $^4$He gas.  
The sample temperature was decreased by cooling the quartz tube with the liquid $^4$He. 

Microwaves with a frequency of $\sim$70 GHz were generated using the microwave oscillator 
(ELVA-1 Ltd., VCOM-12) and guided into the brass box through a waveguide. 
The continuous wave NMR system with a Liverpool Q-meter was used 
to measure the enhancement of the polarization by sweeping 
the RF frequency~\cite{Liverpool}. 
As shown in Fig.~\ref{fig:apparatus}(b-2, c-2), the NMR coils were 
wound on the samples. 
In the case of the sample with Nd 0.05 mol\% (Fig.~\ref{fig:apparatus}(b-2)), 
the NMR RF field was tilted by 20$^{\circ}$ from the perpendicular angle 
of the external magnetic field B$_{0}$. 
It is noted that a reduction of the NMR sensitivity due to the tilted coil is 
estimated to be about 10\%. 

\subsection{NMR measurements}

%\begin{figure}[ht]
%\centering
%%\includegraphics[keepaspectratio, scale=0.35]{samples.jpg}
%\includegraphics[width=8.5cm]{samples.jpg}
%\caption{The NMR coils and LaAlO$_3$ samples with (a) Nd 0.05 mol\% 
%         and (b) Nd 0.01 mol\%. }
%\label{fig:samples}
%\end{figure}

We calculate the NMR peak shift $\Delta \nu_m$ from the peak of the 
$-1/2 \leftrightarrow +1/2$ transition using 
\begin{equation}
  \Delta \nu_m =\nu_{m \leftrightarrow m+1} -\nu_{-1/2 \leftrightarrow +1/2} .
  \label{eq:peak_shift}
\end{equation}
The calculated $\Delta \nu_m$ for $^{27}$Al is listed 
in Table~\ref{table:expected_Delta_nu}. 

\begin{table}[ht]
    \centering
    \caption{NMR peak assignment and peak shift $\Delta \nu_{m \leftrightarrow m+1}$ 
    for $^{27}$Al in the primary (p) and secondary (s) domains at 1.3 K.}
    \begin{tabular}{cccc}
    \hline
    Peak index & $m \leftrightarrow m+1$ & $\Delta \nu_m$ [kHz] & domain \\
    \hline
    (1) & $+3/2 \leftrightarrow +5/2$ & $-$60 & p \\
    (2) & $+1/2 \leftrightarrow +3/2$ & $-$30 & p \\
    (3) & $-5/2 \leftrightarrow -3/2$ & $-$20 & s \\
    (4) & $-3/2 \leftrightarrow -1/2$ & $-$10 & s \\
    (5) & $-1/2 \leftrightarrow +1/2$ & 0 & p+s \\
    (6) & $+1/2 \leftrightarrow +3/2$ & 10 & s \\
    (7) & $+3/2 \leftrightarrow +5/2$ & 20 & s \\
    (8) & $-3/2 \leftrightarrow -1/2$ & 30 & p \\
    (9) & $-5/2 \leftrightarrow -3/2$ & 60 & p \\

    \hline
    \end{tabular}
    \label{table:expected_Delta_nu}
\end{table}

As indicated in Table~\ref{table:expected_Delta_nu}, the two peak intensities of 
the primary and secondary domains are mixed in the transition of 
$-1/2 \leftrightarrow +1/2$ because of the lack of an anisotropic aspect. 
The microwave frequency was tuned to enhance the NMR peaks of the primary domain. 
We used the peak intensities of the $+1/2 \leftrightarrow +3/2$ (peak (2)) 
and $-3/2 \leftrightarrow -1/2$ (peak (8)) transitions to evaluate the 
enhancement of the $^{27}$Al polarization for the samples with Nd 0.05 mol\% 
and 0.01 mol\%, respectively, because the peaks were attributed only to 
the primary domain and were easily identified in the $^{27}$Al-NMR spectrum. 
All NMR peaks were within a sweep range of the RF frequency, $\sim$0.2 MHz, 
which was determined by the Q curve. 
In contrast, only one NMR peak was observed for $^{139}$La 
in a sweep range because of the large quadrupole interaction. 
For evaluating the enhancement of the $^{139}$La polarization, we observed 
the center peak corresponding to the $-1/2 \leftrightarrow +1/2$ transition. 
NMR signals were measured at 1.3 K and 4.2 K, and baseline shifts attributed to 
different cooling condition were not observed. 
The sensitivity of NMR measurements was inferred to be unchanged. 

%Previous DNP experiment has observed a spontaneous recovery of the $^{139}$La polarization 
%in the Nd$^{3+}$:LaAlO$_3$ crystal of Nd 0.03 mol\%, which is explainable in terms of 
%contact to the spin-spin reservoir~\cite{Maekawa1995}. 
%In the simplest model of relaxation via the direct dipole-dipole interaction with 
%electron spins, the relaxation curve of $^{139}$La in LaAlO$_3$ crystal is modeled 
%with multi-exponential function because the nuclear spin of $^{139}$La in the static 
%magnetic field is an eight-level system. 
%On the other hand, in the thermal mixing model, it is known that the relaxation curve 
%cannot be analytically expressed at low temperature, even in a $1/2$ nuclear spin system. 
%However, in the previous DNP experiment with the LaAlO$_3$ crystals~\cite{Maekawa1995}, 
%the experimental data of the relaxation curve was roughly analyzable with the single 
%exponential function. 
%In this study, hence, we approximately estimated time constants 
%by phenomenologically fitting a single exponential function to the relaxation curve 
%and buildup curve.

\subsection{Enhancement of the $^{139}$La and $^{27}$Al polarizations with the Nd 0.05 mol\% sample}\label{sec:Nd005}

\begin{figure}[ht]
  \centering
  \includegraphics[width=8.5cm]{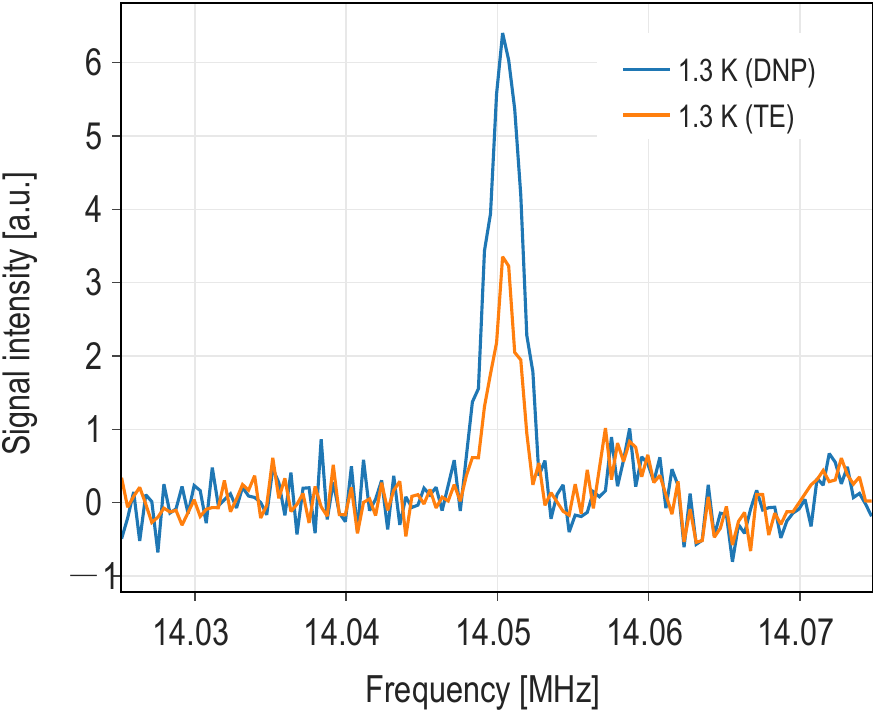}
  \caption{$^{139}$La-NMR spectra of the $-1/2 \leftrightarrow +1/2$ transition 
           with the Nd 0.05 mol\% sample at 1.3 K and 2.3 T.}
  \label{fig:nmr1}
\end{figure}

Figure~\ref{fig:nmr1} shows two $^{139}$La-NMR spectra, where one was measured after 
a sufficiently long time of microwave irradiation, and the other was obtained 
at the thermal equilibrium(TE) state at 1.3 K and 2.3 T. 
The peak of the $-1/2 \leftrightarrow +1/2$ transition is clearly observed and 
a small enhancement of the $^{139}$La polarization is successfully observed. 

\begin{figure}[ht]
  \centering
  \includegraphics[keepaspectratio, scale=0.62]{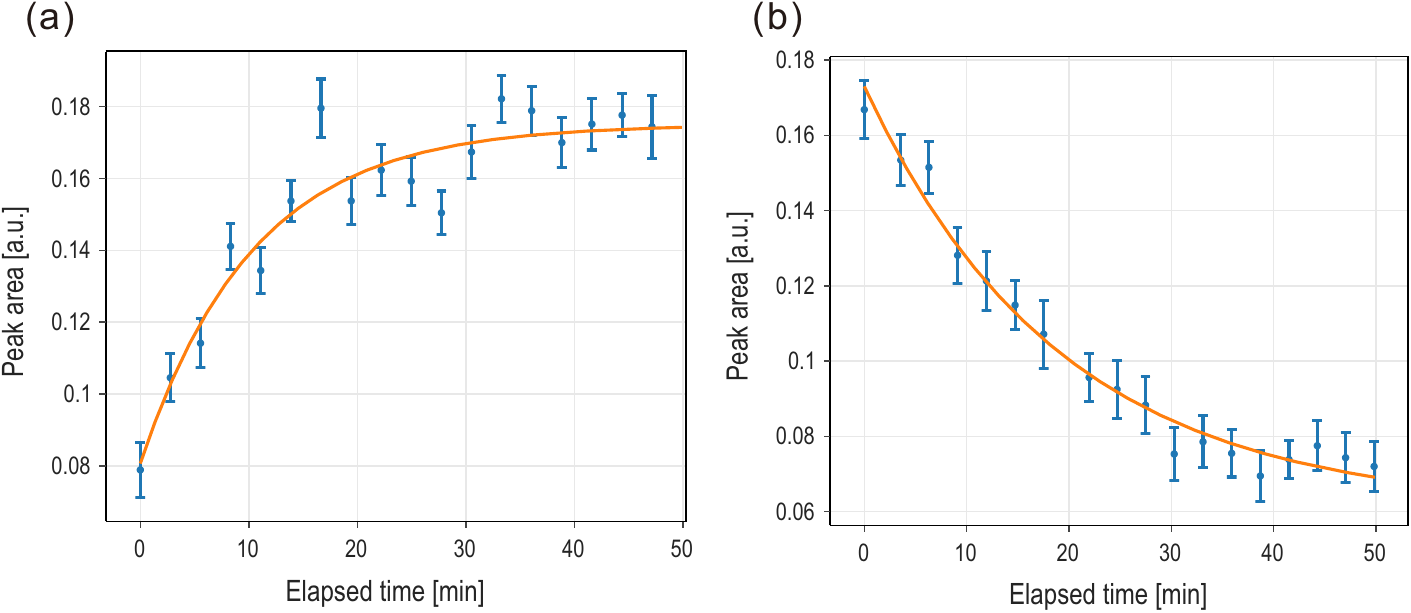}
  \caption{(a) Time dependence of the peak area of the $-1/2 \leftrightarrow +1/2$ transition 
             in $^{139}$La-NMR (Nd 0.05 mol\%) after inputting the 69.345 GHz microwaves.
           (b) Time dependence of the peak area of the $-1/2 \leftrightarrow +1/2$ transition 
             in $^{139}$La-NMR (Nd 0.05 mol\%) after turning off the microwaves. }
  \label{fig:nmr2}
\end{figure}

Figure~\ref{fig:nmr2} (a) shows the build-up curve of the $^{139}$La polarization 
after inputting microwaves with a frequency of 69.345 GHz. 
The build-up time is obtained as $10 \pm 2$ min by fitting the curve with 
an exponential function. 
If the contribution of the secondary domain is ignored, 
the $-1/2 \leftrightarrow +1/2$ signal after inputting the microwaves 
becomes $2.3 \pm 0.2$ times larger than that at the TE state. 
%More precisely, evaluating the contribution only from the primary domain is necessary. 
%We will discuss the analysis considering the secondary domain's contribution later. 

%\begin{figure}[ht]
%    \begin{tabular}{cc}
%        \begin{minipage}{.5\textwidth}
%            \centering
%            \includegraphics[width=1.0\linewidth]{fig5_build.pdf}
%            \caption{Time dependence of the peak area of the $-1/2 \leftrightarrow +1/2$ transition 
%             in $^{139}$La-NMR (Nd 0.05 mol\%) after inputting the 69.345 GHz microwaves.}
%            \label{fig:nmr2_left}
%        \end{minipage}
%        \begin{minipage}{.5\textwidth}
%            \centering
%            \includegraphics[width=1.0\linewidth]{fig5_decay.pdf}
%            \caption{Time dependence of the peak area of the $-1/2 \leftrightarrow +1/2$ transition 
%             in $^{139}$La-NMR (Nd 0.05 mol\%) after turning off the microwaves. }
%            \label{fig:nmr2_right}
%        \end{minipage}
%    \end{tabular}
%\end{figure}

After turning off the microwaves, the NMR peak intensity became smaller and was 
closer to that at the TE state, as shown in Fig.~\ref{fig:nmr2} (b). 
The decay curve of the polarization is fitted with an exponential function, and 
the relaxation time of the polarization is obtained as $19 \pm 2$ min. 
The build-up time is roughly two times shorter than the relaxation time. 
 
\begin{figure}[ht]
\centering
\includegraphics[width=8.5cm]{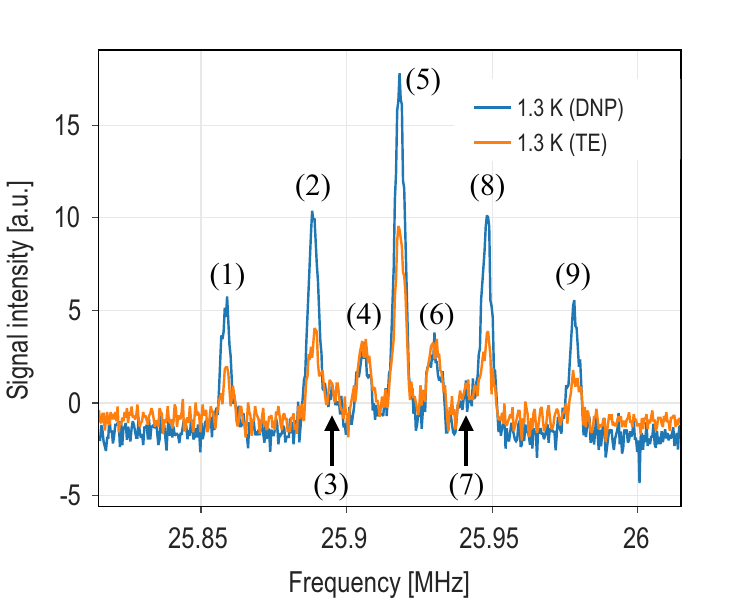}
\caption{$^{27}$Al-NMR spectrum of the sample with Nd 0.05 mol\% at 2.3 T. 
The indices (1), ..., (9) represent peak index referred in 
Table~\ref{table:expected_Delta_nu}.}
\label{fig:nmr3}
\end{figure}

The results of $^{27}$Al under the same condition reveal two NMR spectra at the 
TE and saturated states, as shown in Fig.~\ref{fig:nmr3} . 
Seven peaks are clearly observed at both states, and additional two 
structures, indicated as (3) and (7), are also observed.  
The peak assignment is shown in Table~\ref{table:expected_Delta_nu}. 
The peak intensity of the $+1/2 \leftrightarrow +3/2$ transition becomes 2.5$\pm$0.2 
times larger than that measured at the TE state. 
This enhancement corresponds to an absolute vector polarization of 0.28$\pm$0.02\%. 
The uncertainties are mostly due to the fluctuation of the baselines 
under the NMR peaks. 
Further, only peaks of the primary domain were enhanced by the irradiation of the 
microwaves, and the peaks of the secondary domain were unchanged. 
This implies that the microwave frequency was well tuned to the primary domain. 
The NMR spectrum at the TE state provides the volume information 
of the domains in the sample with Nd 0.05 mol\%. 
The $^{27}$Al-NMR intensity of the peak (6) is nearly equal to that of 
the peak (2) at the TE state in Fig.~\ref{fig:nmr3}. 
These peaks are for the $+1/2 \leftrightarrow +3/2$ transition, and 
the peak (6) and peak (2) are in the secondary and primary domains, respectively, 
as shown in Table~\ref{table:expected_Delta_nu}. 
This is also the case for the (4) and (8) peaks. 
The volumes of the primary and secondary domains are inferred to be nearly equal. 

The results of $^{27}$Al indicate that the presence of the secondary domains 
does not affect 
the enhancement in the primary domain. 
We assume that only the component of the primary domain is 
enhanced regardless of the presence of the secondary domain. 
Although the $-1/2 \leftrightarrow +1/2$ peak at the TE state 
contains the contribution of both primary and secondary domains, the assumption allows 
us to estimate the peak intensity of only the primary domain in the $^{139}$La-NMR spectrum 
at the TE state. 
If equal volumes of the primary and secondary domains are assumed, 
the enhancement for $^{139}$La at the saturated state 
is estimated to be $3.5\pm0.3$ which corresponds an absolute vector polarization 
of 0.27$\pm$0.02\%. 
The uncertainties are mostly due to the fluctuation of the baselines under the 
NMR peaks. 

\subsection{Enhancement of the $^{27}$Al polarization with the Nd 0.01 mol\% sample}\label{sec:Nd001}

The spin temperature of $^{139}$La is estimated to be approximately equal to that of $^{27}$Al 
at all times. 
Considering that the observation of the $^{27}$Al-NMR is considerably easier than 
that of $^{139}$La-NMR, the $^{27}$Al signals are very useful as an indicator for 
comparing two samples with Nd 0.05 and 0.01 mol\%. 

\begin{figure}[ht]
\centering
\includegraphics[width=8.5cm]{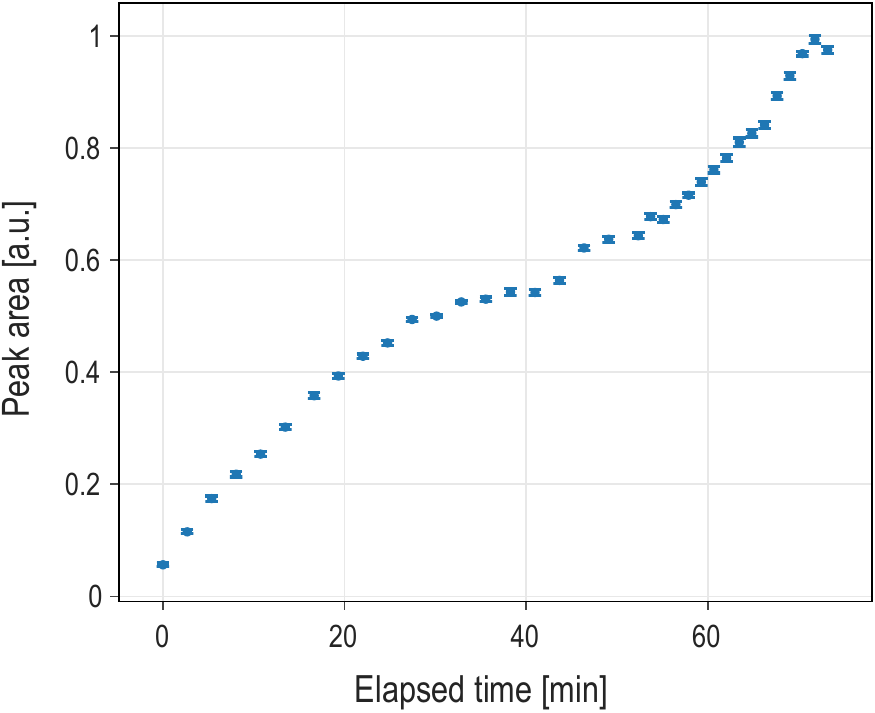}
\caption{
The time dependence of the area of the $^{27}$Al-NMR peak 
(peak (8) in Table~\ref{table:expected_Delta_nu}) 
during the irradiation of microwaves with frequencies of about 69.3 GHz. 
The frequencies were tuned three times between 40 and 70 min. }
\label{fig:buildup_04}
\end{figure}

Figure~\ref{fig:buildup_04} shows that the $^{27}$Al polarization grew 
after inputting microwaves. 
The polarization growth was saturated once between the elapsed times of 30 and 40 min, 
and the growth restarted around 40 min and continued until 70 min. 
The frequencies of the microwaves were changed at 40, 55, 63 min between 
69.345 and 69.360 GHz. 
As a result, the time dependence of the peak area became different from a typical 
build-up curve. 
The measurement time was limited by the amount of liquid $^{4}$He stored 
in the dewar, which was not equipped with a system that could supply  
liquid $^{4}$He continuously. 
Although the polarization growth is not saturated, 
this result implies that the relaxation time of the $^{27}$Al polarization 
at 1.3 K and 2.3 T is longer than 70 min. 
The observation of the TE NMR peaks at 1.3 K was difficult because of the limitation 
of time for leaving the sample. 
Thus, we measured the intensity of the TE signal at 4.2 K with an accuracy of 22\% 
and estimated the intensity at 1.3 K by multiplying with a factor 
of 3.23(=$4.2/1.3$). 
The intensity of the NMR signal after the DNP was measured with an accuracy 
of 5\%. 
Although the $^{27}$Al polarization growth is not saturated, the maximum 
enhancement at 1.3 K results in 13$\pm$3 which corresponds to an absolute vector 
polarization of 1.4$\pm$0.3\%. 
The uncertainties are mostly due to the fluctuation of the baselines under 
the NMR peaks. 
Therefore, the enhancement with the Nd 0.01 mol\% sample is at least 
5 times larger than that with the Nd 0.05 mol\% sample, 
although their experimental setups are not exactly identical.

\section{Summary and future prospects}

We grew two LaAlO$_3$ crystals with Nd 0.05 and 0.01 mol\%, 
and performed the first DNP experiment at 1.3 K and 2.3 T. 
We succeeded in observing the enhancement of NMR signals for $^{139}$La and $^{27}$Al. 
The DNP for positive nuclear polarizations was performed, and 
that for negative polarization was not attempted. 
For the sample with Nd 0.05 mol\%, we observed an enhancement of 
$3.5 \pm 0.3$ for $^{139}$La. 
The result of the 0.01 mol\% sample showed that the enhancement for $^{27}$Al was 
$\sim$5 times larger than that of the 0.05 mol\% sample, although the 
experimental setups were not completely identical. 
These enhancements correspond to absolute vector polarizations 
of 0.27 $\pm$ 0.02\% (Nd 0.05 mol\%) and 1.4 $\pm$ 0.3\% (Nd 0.01 mol\%). 
The experimental results and comparison with 
the previous result of Nd 0.03 mol\% confirm that our method 
can provide samples that are sufficiently effective for DNP.  
Further, a Nd concentration less than 0.03 mol\% was found to be 
suitable for improving the polarization of the polarized target. 

The size of the polarized target for the T-violation experiment 
is a few 10 times larger than that of the present crystal samples. 
However, it is difficult to grow such a large crystal with the optical FZ method. 
One possibility is to apply another method, which can grow larger LaAlO$_3$ 
crystals, with a crucible. 
Contamination deteriorating the performance of the polarized target 
must be removed from the crucible. 
The microwave heat generated in the crystal is accumulated in the large crystal. 
The heat conductivity of the LaAlO$_{3}$ crystal is too small to 
sufficiently cool the inside of the crystal from the surface at a low temperature. 
Dividing the crystal into some pieces can be a good way for the cooling. 
Another possibility is to grow many crystals with the same size as the present ones 
using the optical FZ method, and to combine them as a target. 
The $C_{3}$ axes of all crystals are aligned to the external magnetic 
field, and therefore, special treatments are required for the installation 
and cooling of the crystals. 

The relaxation times of the polarizations were not measured in the 
present experiment because a one-shot cryostat was used. 
The temperature of 1.3 K lasted for only 6 h in which 
the relaxation time could not be measured precisely after the build-up 
of the polarization. 
Currently, we are developing a new cryostat that is operated continuously 
by filling it with liquid helium. 
The new cryostat has a needle valve that produces the pressure difference between 
the sample space and liquid helium bath. 
A relaxation time longer than 1 day will be measured in the near future. 

Toward the realization of the polarized $^{139}$La target, we established 
the primary method for growing LaAlO$_3$ crystals doped with Nd$^{3+}$ ions with 
the optical FZ method using halogen lamps and evaluating the crystals with 
the DNP. 
This is expected to play a critical role in finding the optimal amount of 
Nd$^{3+}$ ions to obtain a higher polarization and a longer relaxation time 
of the polarization. 
In fact, our recent experiment with a Nd$^{3+}$ 0.01 mol\% sample achieved 
a larger enhancement for the $^{139}$La-NMR peak~\cite{Ide2023}. 
Most established polarized targets have been limited to proton and deuteron targets. 
Our positive results in the development of the polarized lanthanum target are unique 
in the nuclear and particle physics field. 
We hope the present method for target development will be used for not only the 
lanthanum targets but also for various nuclear targets in the future. 

%\section{acknowledgments}

\section{AUTHOR DECLARATIONS}
{\bf Conflict of Interest}\\
The authors have no conflicts to disclose. \\

{\bf Data Availability}\\
The data that support the findings of this study are available 
from the corresponding author upon reasonable request. 

\begin{acknowledgments}
    The DNP experiments were performed under the RCNP project "Development of polarized target for new physics search via T-violation" and  RCNP Collaboration Research Network program (Project No. COREnet-026) of the Research Center of Nuclear Physics, Osaka University. The material research was performed under the GIMRT Program of the Institute for Materials Research, Tohoku University (Proposal No. 19K0081, 20K0018, and 20N0002). The EPR experiments were conducted with H. Mino, Nagoya University. 
    We would like to thank Editage (www.editage.jp) for English language editing. 
\end{acknowledgments}
  
  %\bibliographystyle{jpsj}
\bibliography{sample}

\begin{thebibliography}{18}%
\makeatletter
\providecommand \@ifxundefined [1]{%
 \@ifx{#1\undefined}
}%
\providecommand \@ifnum [1]{%
 \ifnum #1\expandafter \@firstoftwo
 \else \expandafter \@secondoftwo
 \fi
}%
\providecommand \@ifx [1]{%
 \ifx #1\expandafter \@firstoftwo
 \else \expandafter \@secondoftwo
 \fi
}%
\providecommand \natexlab [1]{#1}%
\providecommand \enquote  [1]{``#1''}%
\providecommand \bibnamefont  [1]{#1}%
\providecommand \bibfnamefont [1]{#1}%
\providecommand \citenamefont [1]{#1}%
\providecommand \href@noop [0]{\@secondoftwo}%
\providecommand \href [0]{\begingroup \@sanitize@url \@href}%
\providecommand \@href[1]{\@@startlink{#1}\@@href}%
\providecommand \@@href[1]{\endgroup#1\@@endlink}%
\providecommand \@sanitize@url [0]{\catcode `\\12\catcode `\$12\catcode `\&12\catcode `\#12\catcode `\^12\catcode `\_12\catcode `\%12\relax}%
\providecommand \@@startlink[1]{}%
\providecommand \@@endlink[0]{}%
\providecommand \url  [0]{\begingroup\@sanitize@url \@url }%
\providecommand \@url [1]{\endgroup\@href {#1}{\urlprefix }}%
\providecommand \urlprefix  [0]{URL }%
\providecommand \Eprint [0]{\href }%
\providecommand \doibase [0]{http://dx.doi.org/}%
\providecommand \selectlanguage [0]{\@gobble}%
\providecommand \bibinfo  [0]{\@secondoftwo}%
\providecommand \bibfield  [0]{\@secondoftwo}%
\providecommand \translation [1]{[#1]}%
\providecommand \BibitemOpen [0]{}%
\providecommand \bibitemStop [0]{}%
\providecommand \bibitemNoStop [0]{.\EOS\space}%
\providecommand \EOS [0]{\spacefactor3000\relax}%
\providecommand \BibitemShut  [1]{\csname bibitem#1\endcsname}%
\let\auto@bib@innerbib\@empty
%</preamble>
\bibitem [{\citenamefont {Wu}\ \emph {et~al.}(1957)\citenamefont {Wu}, \citenamefont {Ambler}, \citenamefont {Hayward}, \citenamefont {Hoppes},\ and\ \citenamefont {Hudson}}]{Wu1957}%
  \BibitemOpen
  \bibfield  {author} {\bibinfo {author} {\bibfnamefont {C.~S.}\ \bibnamefont {Wu}}, \bibinfo {author} {\bibfnamefont {E.}~\bibnamefont {Ambler}}, \bibinfo {author} {\bibfnamefont {R.~W.}\ \bibnamefont {Hayward}}, \bibinfo {author} {\bibfnamefont {D.~D.}\ \bibnamefont {Hoppes}}, \ and\ \bibinfo {author} {\bibfnamefont {R.~P.}\ \bibnamefont {Hudson}},\ }\bibfield  {title} {\enquote {\bibinfo {title} {Experimental test of parity conservation in beta decay},}\ }\href {\doibase 10.1103/PhysRev.105.1413} {\bibfield  {journal} {\bibinfo  {journal} {Phys. Rev.}\ }\textbf {\bibinfo {volume} {105}},\ \bibinfo {pages} {1413--1415} (\bibinfo {year} {1957})}\BibitemShut {NoStop}%
\bibitem [{\citenamefont {Mitchell}\ \emph {et~al.}(2001)\citenamefont {Mitchell}, \citenamefont {Bowman}, \citenamefont {Penttilä},\ and\ \citenamefont {Sharapov}}]{Mitchell2001}%
  \BibitemOpen
  \bibfield  {author} {\bibinfo {author} {\bibfnamefont {G.}~\bibnamefont {Mitchell}}, \bibinfo {author} {\bibfnamefont {J.}~\bibnamefont {Bowman}}, \bibinfo {author} {\bibfnamefont {S.}~\bibnamefont {Penttilä}}, \ and\ \bibinfo {author} {\bibfnamefont {E.}~\bibnamefont {Sharapov}},\ }\bibfield  {title} {\enquote {\bibinfo {title} {Parity violation in compound nuclei: experimental methods and recent results},}\ }\href {\doibase https://doi.org/10.1016/S0370-1573(01)00016-3} {\bibfield  {journal} {\bibinfo  {journal} {Phys. Rep.}\ }\textbf {\bibinfo {volume} {354}},\ \bibinfo {pages} {157--241} (\bibinfo {year} {2001})}\BibitemShut {NoStop}%
\bibitem [{\citenamefont {Gudkov}(1992)}]{Gudkov1992}%
  \BibitemOpen
  \bibfield  {author} {\bibinfo {author} {\bibfnamefont {V.}~\bibnamefont {Gudkov}},\ }\bibfield  {title} {\enquote {\bibinfo {title} {On $\mathrm{CP}$ violation in nuclear reactions},}\ }\href {\doibase https://doi.org/10.1016/0370-1573(92)90121-F} {\bibfield  {journal} {\bibinfo  {journal} {Phys. Rep.}\ }\textbf {\bibinfo {volume} {212}},\ \bibinfo {pages} {77--105} (\bibinfo {year} {1992})}\BibitemShut {NoStop}%
\bibitem [{\citenamefont {Okudaira}\ \emph {et~al.}(2018)\citenamefont {Okudaira}, \citenamefont {Takada}, \citenamefont {Hirota}, \citenamefont {Kimura}, \citenamefont {Kitaguchi}, \citenamefont {Koga}, \citenamefont {Nagamoto}, \citenamefont {Nakao}, \citenamefont {Okada}, \citenamefont {Sakai}, \citenamefont {Shimizu}, \citenamefont {Yamamoto},\ and\ \citenamefont {Yoshioka}}]{Okudaira2018}%
  \BibitemOpen
  \bibfield  {author} {\bibinfo {author} {\bibfnamefont {T.}~\bibnamefont {Okudaira}}, \bibinfo {author} {\bibfnamefont {S.}~\bibnamefont {Takada}}, \bibinfo {author} {\bibfnamefont {K.}~\bibnamefont {Hirota}}, \bibinfo {author} {\bibfnamefont {A.}~\bibnamefont {Kimura}}, \bibinfo {author} {\bibfnamefont {M.}~\bibnamefont {Kitaguchi}}, \bibinfo {author} {\bibfnamefont {J.}~\bibnamefont {Koga}}, \bibinfo {author} {\bibfnamefont {K.}~\bibnamefont {Nagamoto}}, \bibinfo {author} {\bibfnamefont {T.}~\bibnamefont {Nakao}}, \bibinfo {author} {\bibfnamefont {A.}~\bibnamefont {Okada}}, \bibinfo {author} {\bibfnamefont {K.}~\bibnamefont {Sakai}}, \bibinfo {author} {\bibfnamefont {H.~M.}\ \bibnamefont {Shimizu}}, \bibinfo {author} {\bibfnamefont {T.}~\bibnamefont {Yamamoto}}, \ and\ \bibinfo {author} {\bibfnamefont {T.}~\bibnamefont {Yoshioka}},\ }\bibfield  {title} {\enquote {\bibinfo {title} {Angular distribution of $\ensuremath{\gamma}$ rays from neutron-induced compound states of $^{140}\mathrm{La}$},}\ }\href
  {\doibase 10.1103/PhysRevC.97.034622} {\bibfield  {journal} {\bibinfo  {journal} {Phys. Rev. C}\ }\textbf {\bibinfo {volume} {97}},\ \bibinfo {pages} {034622} (\bibinfo {year} {2018})}\BibitemShut {NoStop}%
\bibitem [{\citenamefont {Yamamoto}\ \emph {et~al.}(2020)\citenamefont {Yamamoto}, \citenamefont {Okudaira}, \citenamefont {Endo}, \citenamefont {Fujioka}, \citenamefont {Hirota}, \citenamefont {Ino}, \citenamefont {Ishizaki}, \citenamefont {Kimura}, \citenamefont {Kitaguchi}, \citenamefont {Koga}, \citenamefont {Makise}, \citenamefont {Niinomi}, \citenamefont {Oku}, \citenamefont {Sakai}, \citenamefont {Shima}, \citenamefont {Shimizu}, \citenamefont {Takada}, \citenamefont {Tani}, \citenamefont {Yoshikawa},\ and\ \citenamefont {Yoshioka}}]{Yamamoto2020}%
  \BibitemOpen
  \bibfield  {author} {\bibinfo {author} {\bibfnamefont {T.}~\bibnamefont {Yamamoto}}, \bibinfo {author} {\bibfnamefont {T.}~\bibnamefont {Okudaira}}, \bibinfo {author} {\bibfnamefont {S.}~\bibnamefont {Endo}}, \bibinfo {author} {\bibfnamefont {H.}~\bibnamefont {Fujioka}}, \bibinfo {author} {\bibfnamefont {K.}~\bibnamefont {Hirota}}, \bibinfo {author} {\bibfnamefont {T.}~\bibnamefont {Ino}}, \bibinfo {author} {\bibfnamefont {K.}~\bibnamefont {Ishizaki}}, \bibinfo {author} {\bibfnamefont {A.}~\bibnamefont {Kimura}}, \bibinfo {author} {\bibfnamefont {M.}~\bibnamefont {Kitaguchi}}, \bibinfo {author} {\bibfnamefont {J.}~\bibnamefont {Koga}}, \bibinfo {author} {\bibfnamefont {S.}~\bibnamefont {Makise}}, \bibinfo {author} {\bibfnamefont {Y.}~\bibnamefont {Niinomi}}, \bibinfo {author} {\bibfnamefont {T.}~\bibnamefont {Oku}}, \bibinfo {author} {\bibfnamefont {K.}~\bibnamefont {Sakai}}, \bibinfo {author} {\bibfnamefont {T.}~\bibnamefont {Shima}}, \bibinfo {author} {\bibfnamefont {H.~M.}\ \bibnamefont {Shimizu}},
  \bibinfo {author} {\bibfnamefont {S.}~\bibnamefont {Takada}}, \bibinfo {author} {\bibfnamefont {Y.}~\bibnamefont {Tani}}, \bibinfo {author} {\bibfnamefont {H.}~\bibnamefont {Yoshikawa}}, \ and\ \bibinfo {author} {\bibfnamefont {T.}~\bibnamefont {Yoshioka}},\ }\bibfield  {title} {\enquote {\bibinfo {title} {{Transverse asymmetry of $\gamma$ rays from neutron-induced compound states of $^{140}$La}},}\ }\href {\doibase 10.1103/PhysRevC.101.064624} {\bibfield  {journal} {\bibinfo  {journal} {Phys. Rev. C}\ }\textbf {\bibinfo {volume} {101}},\ \bibinfo {pages} {064624} (\bibinfo {year} {2020})}\BibitemShut {NoStop}%
\bibitem [{\citenamefont {Takahashi}, \citenamefont {Shimizu},\ and\ \citenamefont {Yabuzaki}(1993)}]{Takahashi1993}%
  \BibitemOpen
  \bibfield  {author} {\bibinfo {author} {\bibfnamefont {Y.}~\bibnamefont {Takahashi}}, \bibinfo {author} {\bibfnamefont {H.}~\bibnamefont {Shimizu}}, \ and\ \bibinfo {author} {\bibfnamefont {T.}~\bibnamefont {Yabuzaki}},\ }\bibfield  {title} {\enquote {\bibinfo {title} {Possible nuclear polarization of $\mathrm{^{139}La}$ in $\mathrm{Nd^{3+}}$:$\mathrm{LaAlO_{3}}$ for the test of time reversal invariance},}\ }\href {\doibase https://doi.org/10.1016/0168-9002(93)91266-P} {\bibfield  {journal} {\bibinfo  {journal} {Nucl. Instrum. Methods Phys. Res., Sect. A}\ }\textbf {\bibinfo {volume} {336}},\ \bibinfo {pages} {583--586} (\bibinfo {year} {1993})}\BibitemShut {NoStop}%
\bibitem [{\citenamefont {Maekawa}\ \emph {et~al.}(1995)\citenamefont {Maekawa}, \citenamefont {Takahashi}, \citenamefont {Shimizu}, \citenamefont {Iinuma}, \citenamefont {Masaike},\ and\ \citenamefont {Yabuzaki}}]{Maekawa1995}%
  \BibitemOpen
  \bibfield  {author} {\bibinfo {author} {\bibfnamefont {T.}~\bibnamefont {Maekawa}}, \bibinfo {author} {\bibfnamefont {Y.}~\bibnamefont {Takahashi}}, \bibinfo {author} {\bibfnamefont {H.~M.}\ \bibnamefont {Shimizu}}, \bibinfo {author} {\bibfnamefont {M.}~\bibnamefont {Iinuma}}, \bibinfo {author} {\bibfnamefont {A.}~\bibnamefont {Masaike}}, \ and\ \bibinfo {author} {\bibfnamefont {T.}~\bibnamefont {Yabuzaki}},\ }\bibfield  {title} {\enquote {\bibinfo {title} {{A large nuclear polarization of $^{139}$La in Nd$^{3+}$: LaAlO$_{3}$ for testing the time reversal invariance}},}\ }\href {\doibase 10.1016/0168-9002(95)00584-6} {\bibfield  {journal} {\bibinfo  {journal} {Nucl. Instrum. Methods Phys. Res., Sect. A}\ }\textbf {\bibinfo {volume} {366}},\ \bibinfo {pages} {115--119} (\bibinfo {year} {1995})}\BibitemShut {NoStop}%
\bibitem [{\citenamefont {Hautle}\ and\ \citenamefont {Iinuma}(2000)}]{Hautle2000}%
  \BibitemOpen
  \bibfield  {author} {\bibinfo {author} {\bibfnamefont {P.}~\bibnamefont {Hautle}}\ and\ \bibinfo {author} {\bibfnamefont {M.}~\bibnamefont {Iinuma}},\ }\bibfield  {title} {\enquote {\bibinfo {title} {Dynamic nuclear polarization in crystals of $\mathrm{Nd^{3+}}$:$\mathrm{LaAlO_{3}}$, a polarized $\mathrm{^{139}La}$ target for a test of time-reversal invariance},}\ }\href {\doibase https://doi.org/10.1016/S0168-9002(99)01054-2} {\bibfield  {journal} {\bibinfo  {journal} {Nucl. Instrum. Methods Phys. Res., Sect. A}\ }\textbf {\bibinfo {volume} {440}},\ \bibinfo {pages} {638--642} (\bibinfo {year} {2000})}\BibitemShut {NoStop}%
\bibitem [{\citenamefont {Homer}\ and\ \citenamefont {Brandle}(1967)}]{Fay1967}%
  \BibitemOpen
  \bibfield  {author} {\bibinfo {author} {\bibfnamefont {F.}~\bibnamefont {Homer}}\ and\ \bibinfo {author} {\bibfnamefont {C.}~\bibnamefont {Brandle}},\ }\bibfield  {title} {\enquote {\bibinfo {title} {Czochralski growth and detwinning of $\mathrm{LaAlO_3}$},}\ }in\ \href {http://doi.wiley.com/10.1002/9780470291252.ch19} {\emph {\bibinfo {booktitle} {Crystal growth : proceedings of an International Conference on Crystal Growth, Boston, 20-24 June 1966}}}\ (\bibinfo  {publisher} {Pergamon Press},\ \bibinfo {address} {Hoboken, NJ, USA},\ \bibinfo {year} {1967})\ pp.\ \bibinfo {pages} {51--55}\BibitemShut {NoStop}%
\bibitem [{\citenamefont {Zeng}\ \emph {et~al.}(2004)\citenamefont {Zeng}, \citenamefont {Zhang}, \citenamefont {Zhao}, \citenamefont {Xu}, \citenamefont {Hang}, \citenamefont {Pang}, \citenamefont {Jie}, \citenamefont {Yan},\ and\ \citenamefont {He}}]{Zeng2004}%
  \BibitemOpen
  \bibfield  {author} {\bibinfo {author} {\bibfnamefont {X.}~\bibnamefont {Zeng}}, \bibinfo {author} {\bibfnamefont {L.}~\bibnamefont {Zhang}}, \bibinfo {author} {\bibfnamefont {G.}~\bibnamefont {Zhao}}, \bibinfo {author} {\bibfnamefont {J.}~\bibnamefont {Xu}}, \bibinfo {author} {\bibfnamefont {Y.}~\bibnamefont {Hang}}, \bibinfo {author} {\bibfnamefont {H.}~\bibnamefont {Pang}}, \bibinfo {author} {\bibfnamefont {M.~Y.}\ \bibnamefont {Jie}}, \bibinfo {author} {\bibfnamefont {C.}~\bibnamefont {Yan}}, \ and\ \bibinfo {author} {\bibfnamefont {X.}~\bibnamefont {He}},\ }\bibfield  {title} {\enquote {\bibinfo {title} {Crystal growth and optical properties of $\mathrm{La}\mathrm{Al}\mathrm{O}_{3}$ and $\mathrm{Ce}$-doped $\mathrm{LaAlO_{3}}$ single crystals},}\ }\href {\doibase 10.1016/j.jcrysgro.2004.07.032} {\bibfield  {journal} {\bibinfo  {journal} {J. Cryst. Growth}\ }\textbf {\bibinfo {volume} {271}},\ \bibinfo {pages} {319--324} (\bibinfo {year} {2004})}\BibitemShut {NoStop}%
\bibitem [{\citenamefont {Fahey}, \citenamefont {Strauss},\ and\ \citenamefont {Anderson}(1993)}]{Fahey1993}%
  \BibitemOpen
  \bibfield  {author} {\bibinfo {author} {\bibfnamefont {R.~E.}\ \bibnamefont {Fahey}}, \bibinfo {author} {\bibfnamefont {A.~J.}\ \bibnamefont {Strauss}}, \ and\ \bibinfo {author} {\bibfnamefont {A.~C.}\ \bibnamefont {Anderson}},\ }\bibfield  {title} {\enquote {\bibinfo {title} {{Vertical gradient-freeze growth of aluminate crystals}},}\ }\href {\doibase 10.1016/S0022-0248(07)80022-4} {\bibfield  {journal} {\bibinfo  {journal} {J. Cryst. Growth}\ }\textbf {\bibinfo {volume} {128}},\ \bibinfo {pages} {672--679} (\bibinfo {year} {1993})}\BibitemShut {NoStop}%
\bibitem [{\citenamefont {Suzuki}\ \emph {et~al.}(2005)\citenamefont {Suzuki}, \citenamefont {Ohsato}, \citenamefont {Kakimoto}, \citenamefont {Shimada}, \citenamefont {Sasaki},\ and\ \citenamefont {Saka}}]{Suzuki2005}%
  \BibitemOpen
  \bibfield  {author} {\bibinfo {author} {\bibfnamefont {S.}~\bibnamefont {Suzuki}}, \bibinfo {author} {\bibfnamefont {H.}~\bibnamefont {Ohsato}}, \bibinfo {author} {\bibfnamefont {K.-I.}\ \bibnamefont {Kakimoto}}, \bibinfo {author} {\bibfnamefont {T.}~\bibnamefont {Shimada}}, \bibinfo {author} {\bibfnamefont {K.}~\bibnamefont {Sasaki}}, \ and\ \bibinfo {author} {\bibfnamefont {H.}~\bibnamefont {Saka}},\ }\bibfield  {title} {\enquote {\bibinfo {title} {Growth of $\mathrm{LaAlO_{3}}$ single crystal by floating zone method and its microwave properties},}\ }in\ \href {\doibase 10.1002/9780470291252.ch19} {\emph {\bibinfo {booktitle} {Advances in Electronic Ceramic Materials: Ceramic Engineering and Science Proceedings, Volume 26, Number 5}}}\ (\bibinfo  {publisher} {John Wiley {\&} Sons, Inc.},\ \bibinfo {address} {Hoboken, NJ, USA},\ \bibinfo {year} {2005})\ pp.\ \bibinfo {pages} {177--184}\BibitemShut {NoStop}%
\bibitem [{\citenamefont {Inagaki}\ \emph {et~al.}(2007)\citenamefont {Inagaki}, \citenamefont {Suzuki}, \citenamefont {Kagomiya}, \citenamefont {Kakimoto}, \citenamefont {Ohsato}, \citenamefont {Sasaki}, \citenamefont {Kuroda},\ and\ \citenamefont {Shimada}}]{Inagaki2007}%
  \BibitemOpen
  \bibfield  {author} {\bibinfo {author} {\bibfnamefont {Y.}~\bibnamefont {Inagaki}}, \bibinfo {author} {\bibfnamefont {S.}~\bibnamefont {Suzuki}}, \bibinfo {author} {\bibfnamefont {I.}~\bibnamefont {Kagomiya}}, \bibinfo {author} {\bibfnamefont {K.}~\bibnamefont {Kakimoto}}, \bibinfo {author} {\bibfnamefont {H.}~\bibnamefont {Ohsato}}, \bibinfo {author} {\bibfnamefont {K.}~\bibnamefont {Sasaki}}, \bibinfo {author} {\bibfnamefont {K.}~\bibnamefont {Kuroda}}, \ and\ \bibinfo {author} {\bibfnamefont {T.}~\bibnamefont {Shimada}},\ }\bibfield  {title} {\enquote {\bibinfo {title} {{Crystal structure and microwave dielectric properties of SrTiO$_{3}$ doped LaAlO$_{3}$ single crystal grown by FZ}},}\ }\href {\doibase 10.1016/j.jeurceramsoc.2006.11.045} {\bibfield  {journal} {\bibinfo  {journal} {Journal of the European Ceramic Society}\ }\textbf {\bibinfo {volume} {27}},\ \bibinfo {pages} {2861--2864} (\bibinfo {year} {2007})}\BibitemShut {NoStop}%
\bibitem [{\citenamefont {Bednorz}\ and\ \citenamefont {Arend}(1984)}]{Bednorz1984}%
  \BibitemOpen
  \bibfield  {author} {\bibinfo {author} {\bibfnamefont {J.~G.}\ \bibnamefont {Bednorz}}\ and\ \bibinfo {author} {\bibfnamefont {H.}~\bibnamefont {Arend}},\ }\bibfield  {title} {\enquote {\bibinfo {title} {A 1 $\mathrm{kW}$ mirror furnace for growth of refractory oxide single crystals by a floating-zone technique},}\ }\href {\doibase 10.1016/0022-0248(84)90064-2} {\bibfield  {journal} {\bibinfo  {journal} {J. Cryst. Growth}\ }\textbf {\bibinfo {volume} {67}},\ \bibinfo {pages} {660--662} (\bibinfo {year} {1984})}\BibitemShut {NoStop}%
\bibitem [{\citenamefont {Ishizaki}\ \emph {et~al.}(2020)\citenamefont {Ishizaki}, \citenamefont {Shimizu}, \citenamefont {Kitaguchi}, \citenamefont {Matsushita}, \citenamefont {Iinuma}, \citenamefont {Kohri}, \citenamefont {Yoshikawa}, \citenamefont {Yosoi}, \citenamefont {Shima}, \citenamefont {Iwata}, \citenamefont {Miyachi}, \citenamefont {Ishimoto}, \citenamefont {Fujita},\ and\ \citenamefont {Ikeda}}]{Ishizaki2020}%
  \BibitemOpen
  \bibfield  {author} {\bibinfo {author} {\bibfnamefont {K.}~\bibnamefont {Ishizaki}}, \bibinfo {author} {\bibfnamefont {H.~M.}\ \bibnamefont {Shimizu}}, \bibinfo {author} {\bibfnamefont {M.}~\bibnamefont {Kitaguchi}}, \bibinfo {author} {\bibfnamefont {T.}~\bibnamefont {Matsushita}}, \bibinfo {author} {\bibfnamefont {M.}~\bibnamefont {Iinuma}}, \bibinfo {author} {\bibfnamefont {H.}~\bibnamefont {Kohri}}, \bibinfo {author} {\bibfnamefont {H.}~\bibnamefont {Yoshikawa}}, \bibinfo {author} {\bibfnamefont {M.}~\bibnamefont {Yosoi}}, \bibinfo {author} {\bibfnamefont {T.}~\bibnamefont {Shima}}, \bibinfo {author} {\bibfnamefont {T.}~\bibnamefont {Iwata}}, \bibinfo {author} {\bibfnamefont {Y.}~\bibnamefont {Miyachi}}, \bibinfo {author} {\bibfnamefont {S.}~\bibnamefont {Ishimoto}}, \bibinfo {author} {\bibfnamefont {M.}~\bibnamefont {Fujita}}, \ and\ \bibinfo {author} {\bibfnamefont {Y.}~\bibnamefont {Ikeda}},\ }\bibfield  {title} {\enquote {\bibinfo {title} {{Polarized Lanthanum Target for the T-violation Search in
  Slow Neutron Transmission}},}\ }\href {\doibase 10.22323/1.379.0061} {\bibfield  {journal} {\bibinfo  {journal} {PoS}\ }\textbf {\bibinfo {volume} {PSTP2019}},\ \bibinfo {pages} {061} (\bibinfo {year} {2020})}\BibitemShut {NoStop}%
\bibitem [{\citenamefont {Hayward}\ \emph {et~al.}(2005)\citenamefont {Hayward}, \citenamefont {Morrison}, \citenamefont {Redfern}, \citenamefont {Salje}, \citenamefont {Scott}, \citenamefont {Knight}, \citenamefont {Tarantino}, \citenamefont {Glazer}, \citenamefont {Shuvaeva}, \citenamefont {Daniel}, \citenamefont {Zhang},\ and\ \citenamefont {Carpenter}}]{Hayward2005}%
  \BibitemOpen
  \bibfield  {author} {\bibinfo {author} {\bibfnamefont {S.~A.}\ \bibnamefont {Hayward}}, \bibinfo {author} {\bibfnamefont {F.~D.}\ \bibnamefont {Morrison}}, \bibinfo {author} {\bibfnamefont {S.~A.~T.}\ \bibnamefont {Redfern}}, \bibinfo {author} {\bibfnamefont {E.~K.~H.}\ \bibnamefont {Salje}}, \bibinfo {author} {\bibfnamefont {J.~F.}\ \bibnamefont {Scott}}, \bibinfo {author} {\bibfnamefont {K.~S.}\ \bibnamefont {Knight}}, \bibinfo {author} {\bibfnamefont {S.}~\bibnamefont {Tarantino}}, \bibinfo {author} {\bibfnamefont {A.~M.}\ \bibnamefont {Glazer}}, \bibinfo {author} {\bibfnamefont {V.}~\bibnamefont {Shuvaeva}}, \bibinfo {author} {\bibfnamefont {P.}~\bibnamefont {Daniel}}, \bibinfo {author} {\bibfnamefont {M.}~\bibnamefont {Zhang}}, \ and\ \bibinfo {author} {\bibfnamefont {M.~A.}\ \bibnamefont {Carpenter}},\ }\bibfield  {title} {\enquote {\bibinfo {title} {Transformation processes in $\mathrm{La}\mathrm{Al}\mathrm{O}_{3}$: Neutron diffraction, dielectric, thermal, optical, and raman studies},}\ }\href
  {\doibase 10.1103/PhysRevB.72.054110} {\bibfield  {journal} {\bibinfo  {journal} {Phys. Rev. B}\ }\textbf {\bibinfo {volume} {72}},\ \bibinfo {pages} {054110} (\bibinfo {year} {2005})}\BibitemShut {NoStop}%
\bibitem [{\citenamefont {Court}\ \emph {et~al.}(1993)\citenamefont {Court}, \citenamefont {Gifford}, \citenamefont {Harrison}, \citenamefont {Heyes},\ and\ \citenamefont {Houlden}}]{Liverpool}%
  \BibitemOpen
  \bibfield  {author} {\bibinfo {author} {\bibfnamefont {G.}~\bibnamefont {Court}}, \bibinfo {author} {\bibfnamefont {D.}~\bibnamefont {Gifford}}, \bibinfo {author} {\bibfnamefont {P.}~\bibnamefont {Harrison}}, \bibinfo {author} {\bibfnamefont {W.}~\bibnamefont {Heyes}}, \ and\ \bibinfo {author} {\bibfnamefont {M.}~\bibnamefont {Houlden}},\ }\bibfield  {title} {\enquote {\bibinfo {title} {A high precision $\mathrm{Q}$-meter for the measurement of proton polarization in polarised targets},}\ }\href {\doibase https://doi.org/10.1016/0168-9002(93)91047-Q} {\bibfield  {journal} {\bibinfo  {journal} {Nucl. Instrum. Methods Phys. Res., Sect. A}\ }\textbf {\bibinfo {volume} {324}},\ \bibinfo {pages} {433--440} (\bibinfo {year} {1993})}\BibitemShut {NoStop}%
\bibitem [{\citenamefont {Ide}\ \emph {et~al.}(2023)\citenamefont {Ide}, \citenamefont {Fujita}, \citenamefont {Hotta}, \citenamefont {Iinuma}, \citenamefont {Ikeda}, \citenamefont {Ito}, \citenamefont {Iwata}, \citenamefont {Kitaguchi}, \citenamefont {Kohri}, \citenamefont {Miura}, \citenamefont {Miyachi}, \citenamefont {Okudaira}, \citenamefont {Shimizu}, \citenamefont {Takanashi},\ and\ \citenamefont {Yosoi}}]{Ide2023}%
  \BibitemOpen
  \bibfield  {author} {\bibinfo {author} {\bibfnamefont {I.}~\bibnamefont {Ide}}, \bibinfo {author} {\bibfnamefont {M.}~\bibnamefont {Fujita}}, \bibinfo {author} {\bibfnamefont {H.}~\bibnamefont {Hotta}}, \bibinfo {author} {\bibfnamefont {M.}~\bibnamefont {Iinuma}}, \bibinfo {author} {\bibfnamefont {Y.}~\bibnamefont {Ikeda}}, \bibinfo {author} {\bibfnamefont {Y.}~\bibnamefont {Ito}}, \bibinfo {author} {\bibfnamefont {T.}~\bibnamefont {Iwata}}, \bibinfo {author} {\bibfnamefont {M.}~\bibnamefont {Kitaguchi}}, \bibinfo {author} {\bibfnamefont {H.}~\bibnamefont {Kohri}}, \bibinfo {author} {\bibfnamefont {D.}~\bibnamefont {Miura}}, \bibinfo {author} {\bibfnamefont {Y.}~\bibnamefont {Miyachi}}, \bibinfo {author} {\bibfnamefont {T.}~\bibnamefont {Okudaira}}, \bibinfo {author} {\bibfnamefont {H.~M.}\ \bibnamefont {Shimizu}}, \bibinfo {author} {\bibfnamefont {Y.}~\bibnamefont {Takanashi}}, \ and\ \bibinfo {author} {\bibfnamefont {M.}~\bibnamefont {Yosoi}},\ }\bibfield  {title} {\enquote {\bibinfo {title} {{Current
  status of polarized La target development for T-violation search with slow neutron}},}\ }\href {\doibase 10.22323/1.433.0038} {\bibfield  {journal} {\bibinfo  {journal} {PoS}\ }\textbf {\bibinfo {volume} {PSTP2022}},\ \bibinfo {pages} {038} (\bibinfo {year} {2023})}\BibitemShut {NoStop}%
\end{thebibliography}%

\end{document}